\documentclass[USenglish,oneside,twocolumn]{article}

\usepackage[utf8]{inputenc}
\usepackage[big]{dgruyter_NEW}
 
\cclogo{\includegraphics{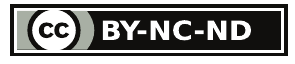}}
  
\usepackage[draft,inline,nomargin]{fixme}   
\usepackage{flushend}
\usepackage{todonotes}
\usepackage{url}
\usepackage{float}
\usepackage{amsmath}
\usepackage{amsfonts}
\usepackage[]{algorithm2e}
\usepackage{algpseudocode}
\usepackage[all]{hypcap}
\makeatletter
\newcommand{\removelatexerror}{\let\@latex@error\@gobble}
\makeatother

\newcommand\myshade{100}
\colorlet{mylinkcolor}{black}
\colorlet{mycitecolor}{black}
\colorlet{myurlcolor}{black}

\hypersetup{
  linkcolor  = mylinkcolor!\myshade!black,
  citecolor  = mycitecolor!\myshade!black,
  urlcolor   = myurlcolor!\myshade!black,
  colorlinks = true,
}

\newlength\figureheight
\newlength\figurewidth
\newlength\smallfigureheight
\newlength\smallfigurewidth
\newlength\largefigureheight
\newlength\largefigurewidth
\newcommand\MarkerSize{2.0}
\newcommand\LineWidth{2.0}
\usepackage{verbatim} 
\usepackage{tikz}
\usepackage{filecontents}
\usepackage{pgfplots, pgfplotstable}
\usetikzlibrary{pgfplots.groupplots,shapes,snakes}
\pgfplotsset{compat=1.5,label style={text height=.5em,text depth=.1em},}
\usepgfplotslibrary{statistics}
\usetikzlibrary{external}

\usepackage{color}
\usepackage{soul}          
\usepackage{balance}
\newcommand{\paragraphX}[1]{\vskip 4pt \noindent \textbf{#1} \hskip .05in}

\newenvironment{smitemize}
  {\begin{list}{--}
     {\setlength{\parsep}{0pt}
      \setlength{\leftmargin}{15pt}
      \setlength{\topsep}{-2pt}
      \setlength{\labelwidth}{5pt}
      \setlength{\itemsep}{1pt}}}
  {\end{list}}

\usepackage[justification=centering]{caption}
\let\oldenumerate\enumerate
\renewcommand{\enumerate}{
  \oldenumerate
  \setlength{\topsep}{0pt}
  \setlength{\leftmargin}{-10pt}
  \setlength{\labelwidth}{10pt}
  \setlength{\itemsep}{2pt}
  \setlength{\parskip}{0pt}
  \setlength{\parsep}{0pt}
}

\usepackage{graphicx}
\usepackage{caption}
\usepackage{subcaption}
\usepackage[numbers]{natbib}

\usepackage{mathtools}
\usepackage{multirow}
\usepackage{color}
\usepackage{soul}          
\usepackage{balance}

\usepackage{graphicx}
\usepackage{caption}
\usepackage{subcaption}
\usepackage[numbers]{natbib}
\usepackage{colortbl}
\usepackage{tikz}
\usepackage{color}
\usepackage{hyperref}
\hypersetup{
    colorlinks=true,
    linkcolor=blue,
    filecolor=magenta,      
    urlcolor=cyan,
}

\newcommand{\attack}{{\em Tik-Tok}}

\begin{document}
  \author*[1]{Mohammad Saidur Rahman}

  \author[2]{Payap Sirinam}

  \author[3]{Nate Mathews}

  \author[4]{Kantha Girish Gangadhara}

  \author[5]{Matthew Wright}

  \affil[1]{Global Cybersecurity Institute, RIT, E-mail: saidur.rahman@mail.rit.edu}

  \affil[2]{Navaminda Kasatriyadhiraj Royal Air Force Academy, E-mail: payap\_siri@rtaf.mi.th}

  \affil[3]{Global Cybersecurity Institute, RIT, E-mail: nate.mathews@mail.rit.edu}

  \affil[4]{Global Cybersecurity Institute, RIT, E-mail: kantha.gangadhara@mail.rit.edu}

  \affil[5]{Global Cybersecurity Institute, RIT, E-mail: matthew.wright@rit.edu}
  
  \title{\huge \attack: The Utility of Packet Timing in Website Fingerprinting Attacks}

  \runningtitle{\attack}


\begin{abstract}
{
A passive local eavesdropper can leverage Website Fingerprinting (WF) to deanonymize the web browsing activity of Tor users. The value of timing information to WF has often been discounted in recent works due to the volatility of low-level timing information. In this paper, we more carefully examine the extent to which packet timing can be used to facilitate WF attacks. We first propose a new set of timing-related features based on \emph{burst-level} characteristics to further identify more ways that timing patterns could be used by classifiers to identify sites. Then we evaluate the effectiveness of both raw timing and \emph{directional timing} which is a combination of raw timing and direction in a deep-learning-based WF attack. Our closed-world evaluation shows that directional timing performs best in most of the settings we explored, achieving: (i) 98.4\% in undefended Tor traffic; (ii) 93.5\% on WTF-PAD traffic, several points higher than when only directional information is used; and (iii) 64.7\% against onion sites, 12\% higher than using only direction. Further evaluations in the open-world setting show small increases in both precision (+2\%) and recall (+6\%) with directional-timing on WTF-PAD traffic. 
To further investigate the value of timing information, we perform an information leakage analysis on our proposed handcrafted features. Our results show that while timing features leak less information than directional features, the information contained in each feature is mutually exclusive to one another and can thus improve the robustness of a classifier.
}
\end{abstract}


	\journalname{Proceedings on Privacy Enhancing Technologies}
	\DOI{Editor to enter DOI}
	\startpage{1}
 	\received{..}
	\revised{..}
	\accepted{..}

	\journalyear{..}
	\journalvolume{..}
	\journalissue{..}

\maketitle

\paragraphX{Resources.}
The code and datasets of this paper are available at: \href{https://github.com/msrocean/Tik\_Tok/}{https://github.com/msrocean/Tik\_Tok/}.
\vspace{-0.6cm}

\section{Introduction}

\begin{figure}[ht]
	\centering
    \vspace{-0.9cm}
    \includegraphics[width=0.95\columnwidth]{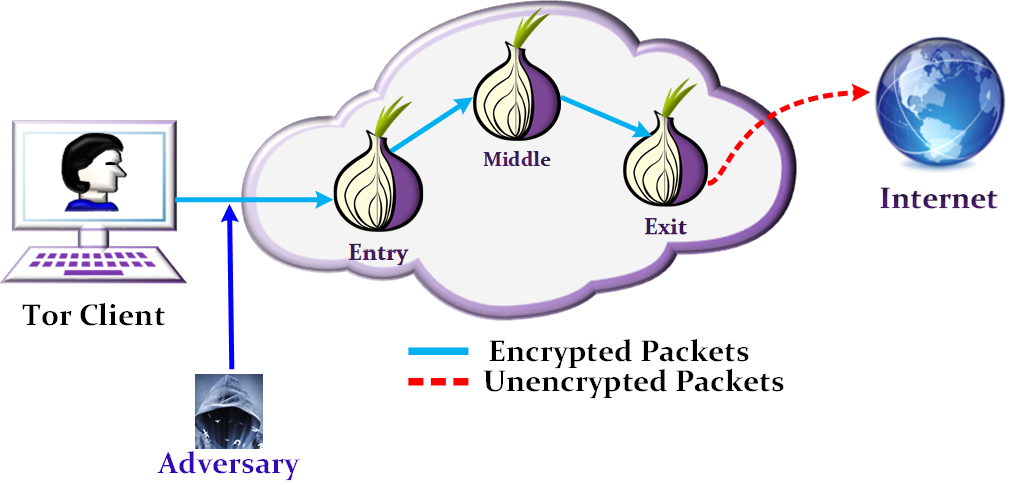}
    \caption{Website fingerprinting threat model.}
    \label{wfattackmodel}
    \vspace{-0.5cm}
\end{figure}


With over eight million daily users~\cite{Mani2018understanding}, Tor is one of the most popular technologies to protect users' online privacy. As shown in Figure~\ref{wfattackmodel}, Tor provides this protection by creating an encrypted \emph{circuit} from the client across three nodes, which relay encrypted traffic between the client and server, typically a website. 
In this design, no single Tor node or eavesdropper should be able to link the user's identity (i.e. IP address and location) with the websites she visits.

Unfortunately, prior work has shown that Tor is prone to a class of traffic analysis attacks called website fingerprinting (WF)~\cite{perry2013critique, wang2014effective, hayes2016k, panchenko2016website, abe2016fingerprinting, Rimmer2018, Sirinam2018, sirinam2019triplet}. The WF attack allows an adversary to learn information about the client's online activities, even though the traffic is encrypted. To perform the attack, a passive local eavesdropper collects side-channel information from the network traffic between the client and entry node, as shown in Figure~\ref{wfattackmodel}. From the collected traffic, the attacker then extracts various features, such as packet statistics or traffic burst patterns, and feeds this information into a machine learning classifier to identify which website the client has visited. Prior work has shown that this kind of attack is very effective, reaching up to 98\% accuracy~\cite{Sirinam2018}.

\begin{figure}[t!]
    \centering
    \includegraphics[scale=0.40]{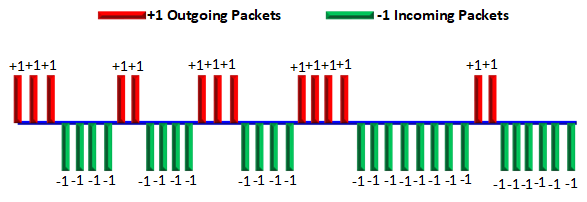}
    \caption{A visualization of bursts, with five outgoing bursts               interspersed with five incoming bursts.}
    \label{burst}
    \vspace{-0.5cm}
\end{figure}

In response, the Tor Project community has become greatly concerned with designing new effective defenses against these WF attacks~\cite{perry2011experimental, perry2013critique}. The state-of-the-art attacks emphasize {\em bursts} as powerful features used to classify the encrypted traces. Bursts are groups of consecutive packets going in the same direction, either \emph{outgoing} from the client to the web server or \emph{incoming} from the server to the client (see Figure~\ref{burst}). Thus, most WF defenses primarily seek to obscure these burst patterns. 

This approach, however, leaves the timing of packets as a largely unprotected source of information for WF attacks to exploit. Moreover, prior work in WF often discounted timing as being not a serious threat~\cite{wang2017walkie}, or found that the contribution of timing was not significant enough when compared to other features to warrant its use~\cite{hayes2016k}. The intuitive explanation for this comes from the fact that timing characteristics are subject to high levels of noise due to many factors, such as varying bandwidth capacity on different circuits. Thus, it appears to be difficult to extract consistent patterns from packet timing that can be used effectively to train WF classifiers. In this work, we investigate new ways timing information can be used in WF attacks, and we find that timing offers significant value to classification.

The key contributions of our work are as follows:
\begin{smitemize}
    \item We develop new \emph{burst-level} timing features and compare them to prior handcrafted features using the WeFDE~\cite{infoleak} information leakage analysis framework. We show that these features are relatively distinct from previously studied features and contribute to the robustness of the classifier.
    
    

    \item We propose the use of a new data representation for the Deep Fingerprinting (DF)~\cite{Sirinam2018} attack and investigate its effects. This new representation, which we refer to as the \emph{Tik-Tok} attack, achieves modest accuracy improvement over direction-only information in several settings. In particular, we reduce the classification error from 9\% to 6.5\% for WTF-PAD by using \emph{Tik-Tok}. Similar performance improvement is seen in the open-world setting for WTF-PAD in which precision and recall are  respectively improved to 0.979 and 0.745 for \emph{Tik-Tok} when tuned for precision. 
    
    \item We perform the first investigation of the performance of deep-learning classifiers on onion services, finding that the DF attack gets only 53\% accuracy, whereas raw timing gets 66\%.

    \item Finally, we develop the first full implementation of the Walkie-Talkie (W-T) defense~\cite{wang2017walkie} in Tor and use it to evaluate our timing-based attacks. We find that our W-T implementation is largely resistant to these attacks despite the fact that the defense was not designed to manipulate timing information. We further discuss our experiences implementing W-T on the live Tor network in the Appendix.

\end{smitemize}
\vspace{0.5cm}
Overall, we find that burst-level timing information can be effectively used as an additional data representation to create an effective WF attack. Moreover, using timing along with packet direction further improves the performance of the attack, especially in the open world. These results indicate that developers of WF defenses need to pay more attention to burst-level timing features as another fingerprintable attribute of users' traffic.


\section{Threat Model}


In this work, we follow a WF threat model that has been frequently applied in the literature~\cite{herrmann2009website,cai2012touching,dyer2012peek,wang2013improved,wang2014effective, panchenko2016website, hayes2016k, Rimmer2018, Sirinam2018, sirinam2019triplet}. We assume that the attacker is a \textit{local} and \textit{passive} network-level adversary. By \emph{local}, we mean that the adversary can be anyone who can access the encrypted streams between the client and the guard, as illustrated in Figure~\ref{wfattackmodel}. This could be an eavesdropper who can intercept the user's wireless connections, the user's Internet service provider (ISP), or any network on the path between the ISP and the guard. By \emph{passive}, we mean limiting the capability of the attacker to only record the encrypted traffic but not delay, drop, or modify it. We also assume that a WF attacker does not have the capability to decrypt the collected encrypted traffic. 

In a WF attack, the adversary needs to first train the WF classifier. To do this, she selects a set of sites that she is interested in classifying and uses Tor to visit these sites a number of times, capturing the network trace of each visit as a sample for that site. From this dataset, she extracts meaningful features and uses them to train the classifier. Once the classifier is trained, she can perform the attack. She intercepts the user's encrypted traffic stream, extracts the same features as used in training, and applies the classifier on those features to predict the user's website. 

Due to the requirement of gathering samples of each site of interest, it is impossible to train the classifier to recognize all possible websites the user might visit. The attacker thus trains the classifier on a limited set of websites called the \emph{monitored set}. All other websites form the \emph{unmonitored set}.


Based on these two sets, researchers have developed two different settings in which to evaluate the performance of the attack:
\emph{closed-world} and \emph{open-world}. In the \emph{closed-world} setting, the user is restricted to visiting only websites in the monitored set. This assumption is generally unrealistic~\cite{Juarez2014, perry2013critique}, but it is useful for evaluating the quality of machine learning (ML) models and potential defenses. In the more realistic \emph{open-world} setting, the user may visit any website, including both monitored and unmonitored sites. This setting is more challenging for the adversary, as she must determine both whether the user is visiting one of the monitored sites and, if so, which one. 
Since it is difficult to produce a dataset covering the entire web, researchers model the open-world setting by using a dataset with samples from many more unmonitored sites than the number of sites in the monitored set. Evaluation in the open-world setting provides more realistic assessments of the effectiveness of both attacks and defenses.

\section{Background and Related Work}



\subsection{WF Attacks using Hand-crafted Features}
\label{wfattack_hancrafted}

Many prior WF attacks apply machine learning (ML) with hand-crafted features. In these attacks, the adversary has to perform \emph{feature engineering}, which is the process of designing a set of effective features that can be used to train the classifier. 

Many WF attacks on HTTPS rely on packet size~\cite{herrmann2009website,miller2014know}, but this is ineffective against Tor, which has fixed-sized cells. Some early WF attacks attempted to use timing information, but with limited success. In 2005, Bissias et al.~\cite{bissias2005privacy} proposed an attack using inter-packet delays averaged over the training set as a profile of that site. The attack is not very effective and was not tested on Tor traffic. In our work, we propose timing features based on bursts of traffic instead of individual packet times.

Panchenko et al. propose an attack with a number of features based on packet volume and packet direction~\cite{panchenko2011website}. They used a support vector machine (SVM) classifier and achieved 55\% accuracy against Tor. Although the paper mentions the use of timing information, none of the features are based on timing, and packet frequency was mentioned as not improving their classification results. Cai et al. proposed an effective (but computationally expensive) attack using no timing information but only the direction of each packet~\cite{cai2012touching}.

WF attacks using hand-crafted features have since been improved using better feature sets and different machine learning algorithms. Four such attacks could attain over 90\% accuracy and have been used as benchmarks for the subsequent research in WF attacks and defenses. 
In the first part of our work, we have compared with these attacks to evaluate the utility of our new timing-based features. 

\paragraphX{$k-$NN.}
Wang et al. propose an attack using a $k$-nearest neighbor ($k$-NN) classifier on a large and varied feature set~\cite{wang2014effective}. In a closed-world setting of 100 websites, they achieved over 90\% accuracy. 
This attack was the first to use a diverse set of features (bursts, packet ordering, concentration of the packets, number of incoming and outgoing packets, etc.) from the traffic information in a WF attack on Tor. A key set of features they identified is based on the pattern of bursts, while the only timing-related feature is total transmission time.

\paragraphX{CUMUL.}
Using a relatively simple feature set based on packet size, packet ordering, and packet direction,
Panchenko et al. propose an attack using the SVM classifier~\cite{panchenko2016website}. This simple feature set, which did not include timing information, proved effective, with 92\% accuracy in the closed-world setting. 

\paragraphX{$k-$FP.}
Hayes et al. propose the $k-$fingerprinting ($k$-FP) attack, which uses random decision forests (RDFs) to rank the features before performing classification with $k$-NN~\cite{hayes2016k}. This attack also achieved over 90\% accuracy in the closed-world setting. Unlike the $k$-NN and CUMUL, their work did study timing features. They found that statistics on packets per second, e.g. the maximum number of packets sent in one second, were moderately helpful features in classifying sites. One such feature ranked ninth among all 150 features, with a fairly large feature importance score of 0.28, while most of the features ranked between 38 and 50 with feature importance scores of 0.07 and below. Statistics on inter-packet delays were also ranked relatively low, between 40-70. In our work, we explore a novel set of timing features based on bursts of traffic instead of fixed time intervals or individual packets. We also use histograms to capture a broader statistical profile than the maximum, minimum, standard deviation, and quartile statistics primarily used by Hayes et al.

\paragraphX{Wfin}
Yan and Kaur recently released a large set of 35,000 features in their {\em Wfin} attack~\cite{Yan2018,Yan2018-tech}. Their study evaluated the significance of features in seven distinct WF scenarios, of which two of these scenarios 
model undefended and defended Tor traffic.
Wfin achieved 96.8\% accuracy against undefended Tor and 95.4\% with Tor-like traffic defended by sending packets at fixed intervals. When the authors investigated the importance ranking of their features, several timing-based features appeared in the top 30 (six timing features between \#11-30). Similar to Hayes et al., their timing features focus mostly on packets per second either across the entire trace or in the first 20 packets.


\subsection{WF Attacks using Deep Learning}
\label{wfattack_DL}
Deep learning (DL) has recently become the default technique in many domains such as image recognition and speech recognition due to its effectiveness~\cite{LeCun2015}. 
Its performance, especially in image recognition~\cite{Krizhevsky2012,Simonyan2015}, has motivated other research communities to adopt DL to improve classification performance in their work. In WF, five works have examined the use of DL classifiers for attacks, of which only one uses timing information.

\paragraphX{SDAE.}
Abe and Goto were the first to explore DL in traffic analysis~\cite{abe2016fingerprinting}. They propose a \textit{Stacked Denoising Autoencoder (SDAE)} for their classifier and a simple input data representation (which we call \emph{direction-only}) composed of a sequence of {\tt 1} for each outgoing packet and {\tt -1} for each incoming packet, ordered according to the traffic trace. After the final packet of a trace, the sequence was padded to a fixed length with {\tt 0}s. In the closed-world setting, using Wang et al.'s dataset (100 instances per site)~\cite{wang2014effective}, they achieved 88\% accuracy. 

\paragraphX{Automated WF.}
Rimmer et al. \cite{Rimmer2018} 
study three DL models -- SDAE, Convolutional Neural Network (CNN), and Long-Short Term Memory (LSTM) -- 
and compare them with CUMUL. The attacks were trained with a very large dataset with 900 websites and 2500 traces for each site, using the direction-only data representation.
The results show that SDAE, CNN and CUMUL all achieve 95-97\% accuracy in the closed-world setting.

\paragraphX{DF.}
Sirinam et al. propose the Deep Fingerprinting (DF) attack, which utilizes a deeper and more sophisticated CNN architecture than the one studied by Rimmer et al.~\cite{Sirinam2018}. They evaluated their model with a dataset containing 95 websites and 1000 traces each, again with the simple direction-only data representation. In the closed-world setting, the DF attack attains 98\% accuracy, which is higher than other state-of-the-art WF attacks. Moreover, they also evaluated the performance of their attack against two lightweight WF defenses, WTF-PAD and Walkie-Talkie. The results show that the DF attack achieves over 90\% accuracy against WTF-PAD, the defense that is the main candidate to be deployed in Tor. On Walkie-Talkie, the attack achieved 49.7\% accuracy and a top-2 accuracy of 98.4\%. 

\paragraphX{Var-CNN.} Simultaneous with the main studies we present in this paper, Bhat et al. propose a novel DL-based attack with architecture more tailored to the WF problem than the DF architecture~\cite{bhat2019var}. Much like our work, they find that using just timing information (raw timestamps) can make for an effective WF attack on its own. To use timing and direction together, they propose an ensemble approach and find it to be very effective, outperforming both DF and their best direction-only and timing-only attacks. We do not directly compare our models with theirs, as we were not able to obtain their code before conducting our study.

\paragraphX{$p$-FP.} Oh et al., much like the work of Rimmer et al., explore several different DL architectures in their WF study~\cite{oh2019pfp}. Although they examine numerous scenarios, such as search query fingerprinting and WF against TLS proxies, they do not use timing information and their classifiers do not outperform DF in most settings. Thus, we do not compare with their work.

\subsection{Onion Sites}

An \emph{onion service}, formerly known as a {\em hidden service}, protects the identity of a website (an \emph{onion site}) or other server by making it accessible only through special Tor connections. A client who has the onion service's \texttt{.onion} URL builds a Tor circuit and requests to Tor to connect to the service. The client's circuit is then linked up with another Tor circuit that leads to the service itself. 
Onion services provide various kinds of functions such as web publishing, messaging, and chat~\cite{onionServices}.

Onion sites may be fingerprinted more readily than regular sites. Kwon et al. show that onion sites' traffic can be discriminated from regular websites with more than 90\% accuracy~\cite{Kwon2015}. Moreover, Hayes and Danezis~\cite{hayes2016k} find that the onion sites can be distinguished from other regular web pages with 85\% true positive rate and only 0.02\% false positive from a dataset of 100,000 sites. Therefore, the adversary can effectively filter out the onion sites' traffic from the rest of the Tor traffic and apply WF attacks to determine which onion site is being visited. Since the number of onion sites is on the order of thousands, much smaller than the number of regular sites, the WF attacker only has to deal with a fairly small open world once she determines that the client is visiting an onion site. Since a smaller open world makes the attack easier, WF attacks on onion sites are more dangerous compared to fingerprinting regular websites. 

In 2017, Overdorf et al. \cite{overdorf2017unique} collected a large dataset of onion sites, consisting of 538 sites with 77 instances each. They evaluated the $k$-NN, CUMUL, and $k$-FP attacks on their dataset and examined the features that are significant for fingerprinting each site. Among timing features, packets per second was helpful for distinguishing between the smallest 10\% of onion sites. Any further discussion of timing features was limited. 

Recently, Jansen et al.~\cite{insidejob} demonstrated how to perform an effective WF attack from middle relays. They attain up to 63\% accuracy with the CUMUL attack in a closed world of 1,000 onion sites.

In our paper, we explore the effectiveness of new burst-level timing features for fingerprinting onion sites. Additionally, we are the first to examine the effectiveness of applying more powerful DL-based attacks, both with and without timing information, to fingerprinting onion sites.
\vskip -0.3cm



\subsection{WF Defenses}
In our work, we explore the effectiveness of WF attacks against the state-of-the-art defenses that have been shown to be effective with low bandwidth and latency overheads, namely \emph{WTF-PAD} and \emph{Walkie-Talkie}. 


\paragraphX{WTF-PAD.}
Juarez et al. proposed WTF-PAD~\cite{juarez2016toward}, an extension of the \emph{adaptive padding} technique that was originally proposed to defend against end-to-end timing attacks~\cite{shmatikov2006timing}. WTF-PAD detects large delays between consecutive bursts and adds dummy packets to fill the gaps. The defense requires 54\% bandwidth overhead, though it does not directly add any delays to real traffic, and can reduce the accuracy of the $k$-NN attack to below 20\%. Sirinam et al., however, show that several other attacks including DF (90\% closed-world accuracy) and $k$-FP (69\%) perform much better against WTF-PAD~\cite{Sirinam2018}. In this paper, we study how timing information can be used to further improve attack performance against WTF-PAD.

\paragraphX{Walkie-Talkie.}
Wang and Goldberg~\cite{wang2017walkie} propose the Walkie-Talkie (W-T) defense, which aims to make two or more websites look exactly the same to an attacker. First, W-T modifies the browser to use \emph{half-duplex} communication, in which the browser requests a single object at a time from the server. This creates a more reliable sequence of bursts compared with typical \emph{full-duplex} communication, in which the browser makes multiple requests and then receives multiple replies. Given each site's expected traffic trace through a half-duplex connection, which is expressed as a sequence of burst sizes, W-T creates a \emph{supersequence} of the two sites -- a sequence of the maximum of the burst sizes from each site. 
Then, when the user visits either site, W-T will add padding packets to make the site's burst sequence match the supersequence.
In theory, this ensures that both sites have the same traffic patterns and cannot be distinguished, guaranteeing a maximum attack accuracy of 50\%. Wang and Goldberg report high effectiveness against attacks, along with reasonable overheads: 31\% bandwidth and 34\% latency. Sirinam et al. also report under 50\% accuracy for DF against W-T~\cite{Sirinam2018}. Both works, however, applied padding in simulation to W-T traces previously collected from a modified Tor client. 

In this paper, we examine the effectiveness of this defense more carefully with the first experiments on a full implementation of W-T including padding. Since W-T does not attempt to defend packet timing information, it is interesting to explore the effectiveness of timing features in attacks on it. Also, based on our experiences building W-T, we report on major challenges in designing and practically deploying W-T.

\paragraphX{Fixed-Rate Padding.}
Another class of WF defenses uses {\em fixed-rate} packet transmission, including BuFLO~\cite{dyer2012peek}, CS-BuFLO~\cite{cai2014cs} and Tamaraw~\cite{cai2014systematic}. In these defenses, packets are sent at the same rate throughout the duration of the connection, which completely hides timing patterns and low-level burst activity. The only remaining information for the WF adversary is the overall size of the page, which is also partially masked. Unsurprisingly, these defenses have proven effective against all known attacks, but also suffer from 
bandwidth and latency overheads range from 100\% to 300\%, which is considered too costly for deployment in Tor. In this paper, we do not evaluate against this class of defenses, assuming that timing information would be of no benefit to the classifier's performance.
\section{Representing Timing Information}
\label{timingfeats}

\subsection{Timing Features}
\label{timing_feats_}

Even though the most effective WF attacks use deep learning to automatically extract features instead of manually craft them~\cite{Rimmer2018}, manually crafted features are still important for \emph{interpretable machine learning}~\cite{du2019techniques}. In WF, finding and evaluating manually designed features can help in understanding why some sites may be especially vulnerable~\cite{overdorf2017unique} and how to design more effective and efficient defenses. We thus explore new timing features in this work.

Prior work has explored using low-level timing features such as inter-packet delay~\cite{bissias2005privacy} and second-by-second packet counts~\cite{hayes2016k}, which are not consistent from one instance of a website to another. Hayes et al. compensate for this by using only high-level aggregate statistics such as the mean or maximum~\cite{hayes2016k}. We propose a novel set of timing features based on traffic bursts. By using bursts, which are seen as important features in WF~\cite{wang2017walkie,hayes2016k}, we capture more consistent and reliable information than the low-level features studied previously. Additionally, we employ histograms to capture a range of statistics on these burst-level timing data, to get more granular information than simple high-level statistics.

\begin{figure}[h]
	\centering
    \includegraphics[scale=0.25]{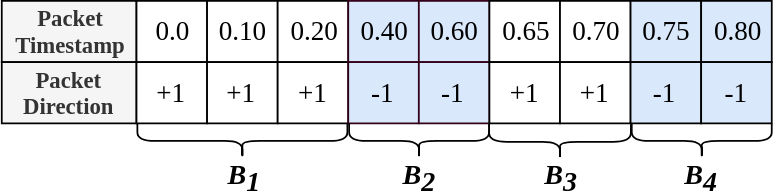}
    \caption{Example of four bursts of traffic ($B_1$, $B_2$, $B_3$, and $B_4$), showing packet directions and timestamps.}
    \vskip -0.3cm
    \label{fourBursts}\vskip -0.8cm
\end{figure}

\subsubsection{Burst-Level Features}
To better describe our burst-level features, we will use a simple example of four bursts shown in Figure~\ref{fourBursts}; outgoing packets are labeled with "$+1$" and incoming with "$-1$". 
Three of our features are focused on the timing of packets inside a single burst. The other five features consider two consecutive bursts.

\paragraphX{Median Packet Time (MED).}
The MED feature represents the median of the timestamps of each burst. For $B_1$ in Figure~\ref{fourBursts}:

\begin{smitemize}
    \item Take the burst's timestamps. $B_1$: [0.0, 0.10, 0.20].
    \item Compute the median. $B_1$: [0.10]. 
\end{smitemize}

\paragraphX{Variance.} 
This feature represents the variance of times within a burst. For $B_1$, we get 0.0067.

\paragraphX{Burst Length.} For this feature, we simply compute the difference between the first and last timestamps in the burst. for $B_1$: $0.20 - 0.0 = 0.2$.


\begin{figure}[t]
	\centering
    \includegraphics[scale=0.225]{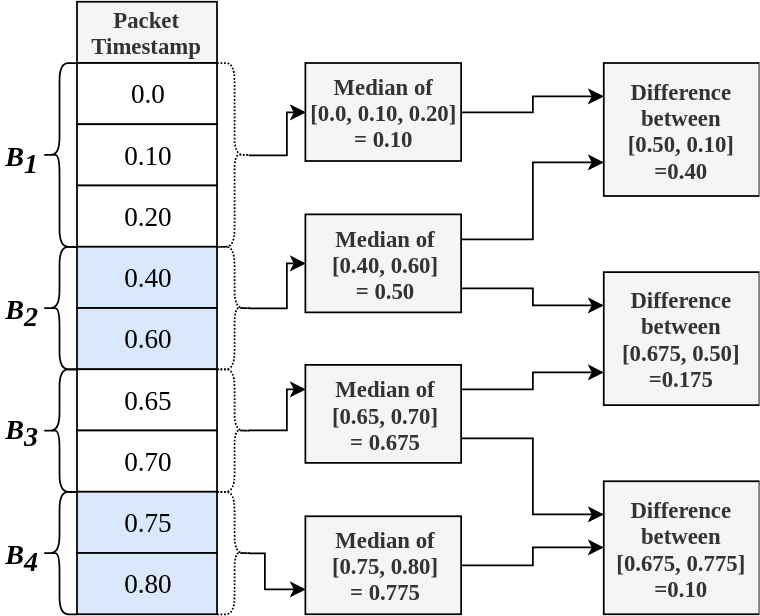}
    \caption{Interval between the Medians of Two Consecutive Bursts: $B_1:B_2$, $B_2:B_3$, and $B_3:B_4$.}
    \label{intermed}
\end{figure}

\paragraphX{IMD.} The extraction process of {\em Inter Median Delay (IMD)}, the interval between the medians of two consecutive bursts, is depicted in Figure~\ref{intermed}. For $B_1$ and $B_2$:

\begin{smitemize}
    \item Compute the medians of each burst as described for computing MED. For $B_1$ and $B_2$, we get 0.10 and 0.50, respectively. 
    \item Take the difference between two consecutive bursts' medians. For $B_1$ and $B_2$: $0.50-0.10=0.40$.
\end{smitemize}

\paragraphX{IBD-FF.}
{\em IBD} stands for {\em Inter-Burst Delay}. {\em IBD-FF} is the interval between the first packets of two consecutive bursts. For $B_1$ and $B_2$: $0.40-0.00=0.40$.


\paragraphX{IBD-LF.}
This feature is the interval between the last packet of one burst and the first packet of the subsequent burst. For $B_1$ and $B_2$, we get $0.40-0.20=0.20$.

\paragraphX{IBD-IFF.}
Similar to IBD-FF, but applied to two consecutive incoming bursts. $B_2$ and $B_4$ are the two incoming bursts in our example, so we get $0.75-0.40=0.35$.


\paragraphX{IBD-OFF.}
Similar to IBD-IFF, but for outgoing bursts. $B_1$ and $B_3$ are the two outgoing bursts in our example, so we get $0.65-0.00=0.65$.


\subsubsection{Histogram Construction}
\label{tuningBins}

To create features that would be robust to changes from instance to instance, we further process the extracted timing features by constructing histograms. 
Just as having quartiles provides more information than just the median, histograms allow us to capture a broader range of statistics for each feature. Our feature processing operates in two phases: (1) produce a global distribution for each feature, and (2) use these global distributions to populate the final feature sets for each instance. 

\begin{table}[t!]
\renewcommand{\arraystretch}{1.25}
  \begin{center}
    \caption{Selecting the number of bins $b$.}\vskip -0.3cm
    \label{tab:binsizes}
    \begin{tabular}{l |c |c}
       \textbf{Dataset} & \textbf{Number of Bins Tested} & \textbf{Selected}\\
      \hline
	Undefended & \multirow{3}{*}{$b$=5$i$ for $i$ $\in$ [1,10]}   & \multirow{4}{*}{$i=4~(b=20)$} \\
        WTF-PAD &           &  \\
        Walkie-Talkie &     &  \\ \cline{1-2}
        Onion Sites & $b$=5$i$ for $i$ $\in$  [1,5] &  \\
    \end{tabular}
  \end{center} 
\end{table}
\paragraphX{Global Distributions.} 
To compute the global distribution for each of our eight features, we begin by computing the raw values of that feature for all instances of every site. We then merge the raw data into a single array, which we sort. This array represents the global distribution for its respective feature. For each feature's global distribution, we then split the data into a histogram with $b$ bins, such that each bin has an equal number of items. The lowest value and highest value in each bin then forms a range for that bin. 

We note that the width of each bin is not constant. For example, considering the MED feature, which represents the median of each burst, there may be many bursts early in the trace. The range for the first bin is thus likely to be quite narrow, going from 0 up to a small value. In contrast, the last bin is likely to have a very wide range. We will use the ranges of the bins when we generate the final feature sets for each instance. In our evaluations, we compute these global distributions separately for our training, validation, and testing datasets so as to accurately model the attacker's capabilities.

\paragraphX{Feature Sets.} Given the histograms created from the global distribution, we generate a set of features for each instance. 
For each of our eight extracted feature sets, we create new histograms of $b$ bins. The range for each bin is given by the bin ranges of the global distribution histograms. The value in each bin is normalized to the range 0 to 1, and this is then a feature used in classification. The full feature set $F$ then includes $|F|=8b$ features, $b$ for each of the eight timing feature types.




\paragraphX{Tuning.}
The number of bins $b$ is a tunable parameter. Using many bins (large $b$) provides more fine-grained features, but it can lead to instability between instances of the same site. Using fewer bins (small $b$) is likely to provide consistent results between instances of the same site, but it does not provide as much detail for distinguishing between sites. We thus explore the variation in classification accuracy for varying values of $b$. We show the search range in Table~\ref{tab:binsizes}. 
Based on our results, we selected $b=20$ for all of our datasets.

\subsection{Raw Timing Information}
To better separate out the importance of timing information from directional information, we investigate the classification performance of using the raw timestamps of all packets in the traffic trace. 
We represent raw timing information as a 1D vector of timestamp values, with a maximum length of 5000, as used for direction-only data in prior work~\cite{Sirinam2018}. An instance with fewer than 5000 packets is padded with zeroes.

\subsection{Combining Timing and Direction Information}
\label{tik_tok_td}


Perhaps the most intuitive way to leverage timing information is to combine it together with directional information and thereby take advantage of both information sources in classification. We call this \emph{directional timing}. Since direction-only represents outgoing packets with $+1$ and incoming packets with $-1$, we propose a directional timing representation generated by simply multiplying the time stamp of each packet by its directional representation. We call the use of directional timing in the DF classifier as the ``\attack~attack.''


\section{Datasets}

\subsection{Undefended, WTF-PAD, Walkie-Talkie, \& the Onion Sites}

For undefended and defended (WTF-PAD \& W-T) traffic, we use the datasets developed by Sirinam et al.~\cite{Sirinam2018}. For Onion Sites, we use the dataset developed by Overdorf et al.~\cite{overdorf2017unique}. The number of sites and the number of instances of each dataset are shown in Table~\ref{tab:datasets}. 

\begin{table}[h]
\renewcommand{\arraystretch}{1.15}
  \begin{center}\vskip -0.5cm
    \caption{Number of Classes and Instances in each Dataset.}\vskip -0.2cm
    \label{tab:datasets}
    \begin{tabular}{l c  c c}
        \textbf{Dataset} & \textbf{Classes}& \textbf{Instances/Class} & \textbf{Total}\\
        \hline 
        Undefended~\cite{Sirinam2018} & 95 & 1000 & 95,000 \\
        WTF-PAD~\cite{Sirinam2018} & 95 &  1000 & 95,000 \\
        Walkie-Talkie (sim.)~\cite{Sirinam2018} & 100 &  900 & 90,000 \\
        Walkie-Talkie (real) & 100 & 750 & 75,000 \\
       Onion Sites~\cite{overdorf2017unique} & 538 & 77 & 41,426 \\
       \hline 
    \end{tabular}\vskip -1.2cm
  \end{center} 
\end{table}

\subsection{Real-world Walkie-Talkie}
\label{sec:real-wt-dataset}

\begin{figure}[t]
	\centering
    \includegraphics[scale=0.25]{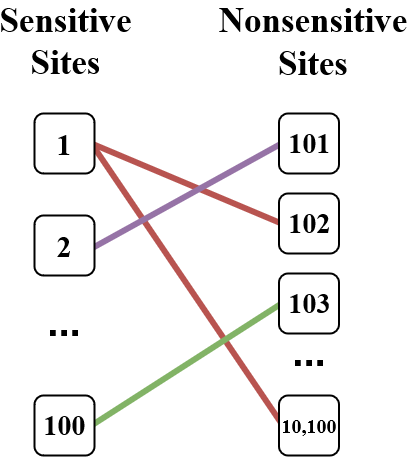}
    \caption{Real-world W-T Website Pairing Strategy.}
    \label{pair}
    \vskip -0.3cm
\end{figure}

The way in which fake timestamps are generated for the padding packets in the simulated W-T dataset is essentially arbitrary and unlikely to be representative of real-world behavior. Consequently, the performance of timing-based attacks on this dataset is not likely to be accurate. To address this, we created the first W-T dataset collected over the live Tor network. We implemented the W-T burst molding algorithm as a Pluggable Transport running on private Tor bridges. We collected a dataset containing 100 sensitive sites paired randomly with 10,000 nonsensitive sites (see Figure~\ref{pair}). Each instance of a site represents one such pairing. 

Unlike prior W-T datasets, our sensitive sites are \emph{not} statically paired with only one nonsensitive site. This mimics a realistic attack scenario, since in the real world, each user would have different pairings of sites, and the attacker would not know which pairings a given user is applying. As a consequence of this collection scheme, one half of the dataset (32,500) represents instances of the client visiting a sensitive site paired with a nonsensitive site, while the other half represents the reverse pairing (a nonsensitive site paired with a sensitive one). 
Additional details of our W-T prototype, as well as discussion of our experiences with W-T, are presented in the Appendix. 




\section{Experimental Evaluation}
\label{SecExperiments}



\subsection{Model Selection}
\label{modelselection}

\paragraphX{Models for Manually Defined Features.}
To select an effective model to study our proposed timing features, we performed experimental evaluations with models used in previous WF attacks: $k$-NN~\cite{wang2014effective} and SVM~\cite{panchenko2016website}. We show the performance of our custom timing features in Section~\ref{subsec:combiningtiming}.

In addition, we use the state-of-the-art Deep Fingerprinting (DF) model~\cite{Sirinam2018} for our experiments with direction-only, our timing features (Section~\ref{subsec:combiningtiming}), raw timing information (Section~\ref{rawtimingexps}), and directional timing (Section~\ref{directionaltiming}). 

\paragraphX{DF Model Architecture.}
The DF model has eight convolutional layers and three dense layers. The last dense layer is the classification layer that returns the probability of each class using softmax regression. Batch normalization is used as the regularizer for both the convolutional layers and the first two dense layers. The model applies max pooling and a dropout rate of 0.1 after each of the two convolutional layers. The first two dense layers also use dropout, with respective dropout rates of 0.70 and 0.50. The model uses both exponential linear unit (ELU) and ReLU activation functions. ReLU is used in most of the layers, but since directional information includes many negative values, ELU is used for the first two convolutional layers. The default number of epochs of the model is 30. For more details, please refer the original paper by Sirinam et al.~\cite{Sirinam2018}. 

For experiments with direction-only data, we reproduced the results reported by Sirinam et al.~\cite{Sirinam2018}, keeping all the hyperparameters the same. For experiments with timing features or directional timing, we do not change any hyperparameters. When training, however, we increase the number of epochs from 30 to 100 for the experiments with timing features and for all experiments on onion services. With timing features, there are 160 features per site, which requires more training to find patterns effectively. For the Onion Sites dataset, there are only 77 instances per site, so more epochs are needed to get the same amount of training.



For the experiments with raw timing data, we made three changes to the hyperparameters: (i) We reduced the dropout rate to 0.40 for both dense layers, which increased the model's performance; (ii) we use the ReLU activation function for the first two convolutional layers instead of ELU, since ReLU is more effective and raw timing includes no negative values; and (iii) we do not use batch normalization in any of the layers.

\subsection{Splitting Data based on Circuits}
\label{split_circuits}

In studying timing characteristics in WF, it is important to model an attacker accurately. One aspect of this is that the attacker should not have access to the same circuit as the victim for generating training data. We model this by ensuring that the attacker's training data is gathered over different circuits from the testing data, as well as the validation data.  

For the Undefended, W-T simulated, and WTF-PAD datasets, Sirinam et al. explained that they collected the datasets in batches within which each website was visited 25 times. The crawler they use is designed to rebuild Tor circuits at the start of each batch. Thus the number of batches used to collect the dataset corresponds to the number of different circuits on which that data was collected. The crawler is also configured to use different entry nodes for each circuit. The Undefended and WTF-PAD datasets have 95 sites with 1000 instances each and were collected over 40 batches. W-T simulated datasets have 100 classes with 900 instances each and were collected over 36 batches.

To correctly model the variance in timing information, we split our dataset such that the circuits used in the training, testing, and validation sets are mutually exclusive. 
We split datasets in a 8:1:1 ratio. For the Undefended and WTF-PAD datasets, instances from the first 32 circuits 
are for training, instances from the next four circuits 
are for validation, and instances from the remaining four circuits 
are for testing. For W-T simulated datasets, instances from the first 29 circuits are used for training, and instances from the next 3 and 4 circuits are used for validation and testing, respectively. For the onion sites dataset, Overdorf et al.~\cite{overdorf2017unique} mention that their crawler creates a new circuit for each visit to crawl their dataset. To collect the W-T real-world dataset, we similarly use our W-T prototype (Appendix~\ref{sec:wtdata}) and create a new circuit for every visit. Thus, for both of these datasets, it is not necessary to manually split the dataset based on the circuits. For both datasets, we use the same 8:1:1 ratio to split into training, validation, and testing sets.




\subsection{Classification Value of Burst-Level Timing Features}
\label{subsec:combiningtiming}


We selected and extracted eight types of burst-level timing features, as 
explained in Section~\ref{timing_feats_}.


We first evaluate our eight features on the Undefended dataset. Table~\ref{tab:undefendedTor} shows our evaluation results for the $k$-NN~\cite{wang2014effective} and SVM~\cite{panchenko2016website} models when used with our features. We also examine the performance of each feature separately to identify the most effective features.
Our initial findings demonstrate that the most effective combination is three features -- MED, IMD, and Burst Length -- when used with $b=20$ bins.
Combining these features together, the accuracy reaches 60.7\% with SVM on Undefended Tor traffic. 

\begin{table}[t]
\renewcommand{\arraystretch}{1.15}
  \begin{center}
    \caption{\textit{Closed-World}: Attack accuracy of burst-level timing features with $k$-NN and SVM against the undefended dataset.}\vskip -0.05cm
    \label{tab:undefendedTor}
    \begin{tabular}{l c c c c}
      &  \multicolumn{4}{c}{\textbf{Features}}\\ \cline{2-5}
      \textbf{Classifier} & MED & IMD & Burst Length & \textbf{Combined} \\
        \hline 
      $k$-NN~\cite{wang2014effective} & 40.6\% & 20.1\% & 18.2\% & 50.8\% \\
      SVM~\cite{panchenko2016website} & 56.3\% & 24.6\% & 29.1\% & \textbf{60.7\%} \\
 		\hline
    \end{tabular}\vskip -0.6cm
  \end{center}
\end{table}

We next apply the combination of all our eight features with a deep-learning model. In particular, we adopt the DF model proposed by Sirinam et al~\cite{Sirinam2018}.
We report the results of each of our closed-world experiments in Table~\ref{tab:CWresults}, where it is compared with our other deep-learning attacks. Note that we do not provide error ranges, as we do not expect that only using our timing features would be a competitive attack.
Nevertheless, the results show that the attack is fairly effective, attaining 84.3\% accuracy against Undefended traffic in the closed-world setting. Against the WTF-PAD and W-T defenses, the attack achieves accuracies of 56.1\% and 80.8\%, respectively. In other words, our features capture much of the valuable information that can be found by a deep-learning classifier on raw data.
Against onion sites, however, the features get just 12.9\% accuracy. We believe that the relatively low number of training examples and the large number of monitored websites are mainly responsible for this result.

\subsection{Information Leakage Analysis}
\label{infor_leakage}

\begin{figure*}[!t]
  \includegraphics[width=\textwidth]{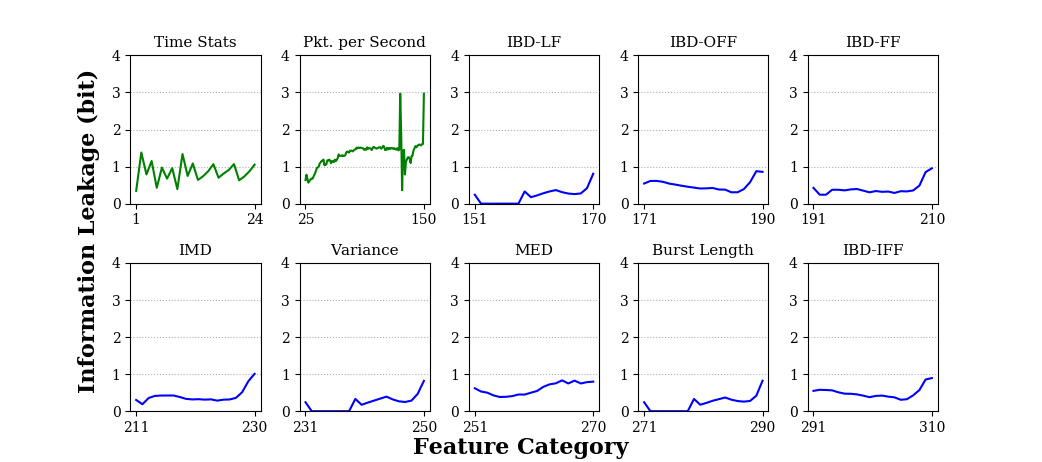}
  \caption{Information leakage for individual features.}
  \label{fig:infoleak}
\end{figure*}


\begin{figure}[!t]
\centering
    \vskip -0.7cm
  \includegraphics[scale=0.75]{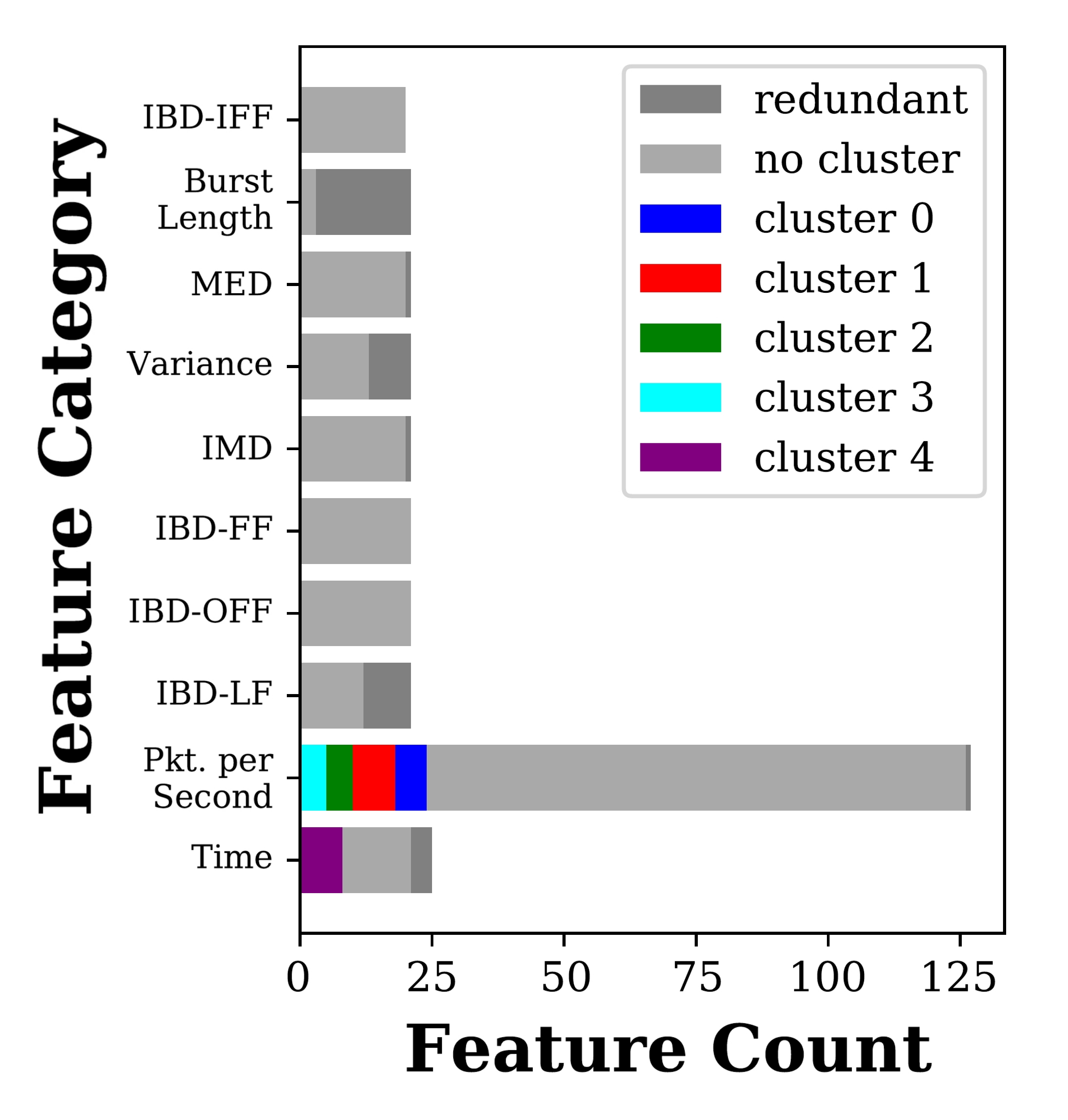}
  \vskip -0.4cm
  \caption{Feature clustering analysis.}
  \label{fig:feat_cluster}\vskip -0.4cm
\end{figure}

To gain a better understanding of the value of burst-level timing features, we adopt the WeFDE~\cite{infoleak} technique and perform an information leakage analysis on the Undefended dataset. This technique allows us to estimate the Shannon bits of information that are revealed by particular features. This type of analysis is appealing, as classification accuracy can fail to capture nuance in feature significance. Although the code of WeFDE paper is publicly available, for speed and memory requirements, we re-implemented most of the code ourselves and validated our common results against theirs to ensure consistency.

WeFDE consists of two components: an \textit{information leakage analyzer} and a \textit{mutual information analyzer}. The information leakage analyzer uses kernel density estimators to model the distributions of features and produce an estimate of the information leakage due to each feature. The mutual information analyzer is instead responsible for producing metrics describing the amount of shared information, i.e. redundancy, that features have with one another. The mutual information analyzer is used to reduce the number of features for leakage analysis by i) identifying features that share most of their information with other features and are thus redundant, and ii) clustering features that show moderate levels of redundancy. For further details, please refer to the paper by Li et al.~\cite{infoleak}.


\begin{table}[tbp]
\renewcommand{\arraystretch}{1.2}
  \begin{center}
      \caption{Joint information leakage of Undefended and WTF-PAD datasets. (* represents our features).}\vskip -0.2cm
    \label{tab:jointleak}
    \begin{tabular}{l c c}
    \multirow{2}{*}{\textbf{Feature Category}} & \multicolumn{2}{c}{\textbf{Leakage (bits)}} \\ \cline{2-3}
    & Undefended & WTF-PAD \\
    \hline 
        \textbf{Pkt. per Second} & 6.56 & 6.56\\
        \textbf{Time Statistics} & 5.92 & 4.68\\
        \textbf{MED*} & 5.43 & 4.75\\
        \textbf{IBD-OFF*} & 4.38 & 3.68\\
        \textbf{IBD-IFF*} & 4.28 & 3.71\\
        \textbf{IBD-FF*} & 3.88 & 3.51\\
        \textbf{IMD*} & 3.87 & 3.45\\
        \textbf{Variance*} & 3.30 & 1.69\\
        \textbf{Burst Length*} & 3.22 & 1.66\\
        \textbf{IBD-LF*} & 3.13 & 1.66\\

    \hline 
    \end{tabular}
  \end{center}
\end{table}

\begin{table*}[th!]
\renewcommand{\arraystretch}{1.25}
  \begin{center}\vskip -0.3cm
      \caption{{\em Closed World:} Comparison of our hand-crafted Timing Features with $k$-FP Timing Features. }
      \vskip -0.2cm
    \label{tab:cw_k_fp}
    \begin{tabular}{l c c c | c c c}
    
    \multicolumn{1}{c}{\multirow{2}{*}{\textbf{Dataset}}}           & \multicolumn{3}{c|}{\textbf{Random Forest}} & \multicolumn{3}{c}{\textbf{Deep Fingerprinting}} \\ \cline{2-7}
    
     & \textbf{$k$-FP}~\cite{hayes2016k} & \textbf{Our}~(\S~\ref{timing_feats_}) & \textbf{$k$-FP + Our} &  \textbf{$k$-FP}~\cite{hayes2016k} & \textbf{Our}~(\S~\ref{timing_feats_}) & \textbf{$k$-FP + Our}\\
    \hline 
		\textbf{Undefended} & 87.4\% & 69.4\%  & 87.3\%  & 89.4\% & 84.3\% & 91.4\% \\
        \textbf{WTF-PAD}  & 69.3\% & 42.1\%  & 69.5\%  & 74.0\% & 56.1\% & 74.0\% \\
        \textbf{Walkie-Talkie (sim.)}  & 76.8\% & 70.0\%  & 80.7\%  & 80.3\% & 80.8\% & 80.5\% \\
        \textbf{Onion Sites}  & 36.3\% &  20.0\%  & 35.5\%  & 33.0\% & 12.8\% & 33.6\% \\
    \hline 
    \end{tabular}\vskip -0.3cm
  \end{center}
\end{table*}


\paragraphX{Results.} First, we perform an analysis of the leakage of individual features, with results shown in Figure~\ref{fig:infoleak}. We find that features in the Packets per Second category appear most significant. The highest amount of leakage from any one feature was 2.97 bits. The distribution of leakage values can be summarized as: only 5\% of features leaked more than 1.52 bits of information, 75\% of features leaked less than 1.30 bits, and 50\% of features leaked no more than 0.69 bits. In general, the information leakage of most individual timing features is low.

Next, we use both components of WeFDE to estimate the information leakage of each feature category, which are depicted in Table~\ref{tab:jointleak}. For this experiment, we use the same values for the clustering threshold (0.40) and redundancy threshold (0.90) as used by Li et al~\cite{infoleak}. 
When calculating the redundancy value of features pairwise, we found that most features have little shared information, i.e. they have low redundancy. 
Because of this, the majority of timing features do not fall into a cluster and are thus modeled as independent variables during the information leakage analysis. The results of our feature redundancy and clustering analysis is shown in Figure~\ref{fig:feat_cluster}. 
We found that just 40 of the 310 timing features are redundant with at least one other feature. Redundant features most often belonged to the Burst Length, Variance, and IBD-LF categories. In addition, only 32 of the 270 non-redundant features could be clustered into five groups. The clusters primarily formed in the previously defined timing feature categories; none of our feature categories formed clusters.

\begin{table*}[tbp]
\renewcommand{\arraystretch}{1.25}
  \begin{center}
      \caption{{\em Closed World:} Accuracy for each attack in different datasets. Error ranges represent standard deviation.}\vskip -0.2cm 
    \label{tab:CWresults}
    \begin{tabular}{l c c c c}
    \textbf{Dataset} & \textbf{Direction}~\cite{Sirinam2018} & \textbf{Timing Features} (\S~\ref{subsec:combiningtiming}) & \textbf{Raw Timing} (\S~\ref{rawtimingexps}) & \textbf{Directional Time} (\S~\ref{directionaltiming})\\
    \hline 
        
        \textbf{Undefended} & \textbf{98.4$\pm$0.1\%} & 84.3\% & $96.5\pm0.3\%$ & \textbf{98.4$\pm$0.1\%} \\
        \textbf{WTF-PAD} & $91.0\pm0.2\%$ & 56.1\%  & $85.9\pm0.6\%$  & \textbf{93.5$\pm$0.7\%} \\
        \textbf{Walkie-Talkie (sim.)} &  47.6$\pm$0.5\% & 80.8\% & $73\pm20\%$  & \textbf{97.0$\pm$0.2\%} \\
        \textbf{Onion Sites} &  $53\pm1\%$ & 12.9\% & \textbf{66$\pm$1\%} & \textbf{64.7$\pm$0.9\%} \\
    \hline 
    \end{tabular}
  \end{center} 
\end{table*}


As a result of this clustering behavior, we find that the joint leakage estimates for each category are significantly higher than what the individual leakage results would lead us to anticipate. We find that the category that leaks the most information is Packets per Second at 6.56 bits. The highest leakage for any of our new features is MED at 5.43 bits. Overall, the low redundancy and consequently higher combined leakage of timing features is a good indicator that even minor features add value to the robustness of the classifier.


\subsection{Comparison of Timing Features with Prior Works}
\label{k_fp_feats_our_feats_compare}
Table~\ref{tab:cw_k_fp} shows a comparison between the performance of our features and $k$-FP~\cite{hayes2016k} timing features on our datasets. We also examine the features' performance with two classifiers: the $k$-FP Random Forest classifier and the DF model. Our Random Forest (RF) classifier uses the same parameters as the $k$-FP RF classifier, but with adjusted feature sets. The DF model follows the design described in Section~\ref{SecExperiments}.

We see that our features provide less classification value than the packets-per-second and other timing features of $k$-FP. 
When our features are combined with $k$-FP features and used in the RF classifier, we see no noticeable improvement. However, when the combined features are instead used as inputs for the DF model, we see a small accuracy improvement of 2\% for the undefended dataset, indicating that DL is slightly more effective at leveraging the additional information provided by our features. This reinforces the results from Section~\ref{infor_leakage}, which showed that our features capture different, but related timing information from features studied in prior works.


\subsection{Raw Timing Information}
\label{rawtimingexps}

From the previous section, we see that using a combination of timing features enables  a reasonably effective WF attack using traditional ML and especially when using DL. It is well known that one of the major advantages of using DL is \textit{end-to-end learning}, in which the classifier can directly learn from the raw input, and this has been shown to provide better performance compared to traditional ML with hand-crafted features~\cite{Song2015end}. Thus, we explore how WF attacks, especially with DL, could effectively perform the attacks by using only raw packet timing. 



\begin{table*}[th!]
\renewcommand{\arraystretch}{1.25}
  \begin{center}\vskip -0.3cm  
\caption{\emph{Real-world W-T:} Accuracy for each attack against the real-world Walkie-Talkie dataset.\\
\label{tab:wt-real}
\textbf{Combined}: both monitored and unmonitored sites included in the test dataset. \\
\textbf{Monitored} \& \textbf{Unmonitored}: the test set includes only the respective instances.}
\begin{tabular}{l c c c}
\textbf{Testing Data}         & \textbf{Direction}~\cite{Sirinam2018} & \textbf{Raw Timing} (\S~\ref{rawtimingexps}) & \textbf{Directional Time} (\S~\ref{directionaltiming})\\ \hline

\textbf{Combined}      & $73.20\pm11.9$\%     & $59.04\pm10.3$\%      & $73.33\pm11.7$\%            \\
\textbf{Monitored}    & $40.53\pm31.7$\%     & $44.92\pm26.2$\%      & $41.59\pm32.2$\%            \\
\textbf{Unmonitored} & $95.08\pm2.5$\%      & $71.97\pm21.2$\%      & $94.28\pm3.44$\%            \\  \hline
\end{tabular}
\end{center}
\end{table*}


In our experiments, we extracted the raw timing information from our datasets and fed them to train a WF classifier using the DF Model. We use stratified $k$-fold cross validation with $k=10$ to obtain standard deviations for better comparison between attacks. Against the Undefended dataset, the attack attains 96\% accuracy, while against the WTF-PAD dataset, it reaches 86\%. For simulated W-T, the attack reaches 73\%, but with a very high variance. 
Most interestingly, on the Onion Sites dataset, we get 13\% higher accuracy using only timing information compared with using only direction information (see~Table~\ref{tab:CWresults}).

As with DL in other domains, WF attacks using DL trained with only timing information have better accuracies compared to our hand-crafted timing features. In the Undefended and WTF-PAD datasets, the attack's respective accuracies improved by 10-25\%. For Onion Sites, we find over 50\% improvement. Overall, our results suggest several takeaways:
~\begin{smitemize}
    \item Using end-to-end learning with the only timing data, the WF classifier can effectively and directly learn more information from the input than the timing features we propose, leading to higher classification performance. 
    \item Even if the only timing data is noisy, the timing information leaves fingerprintable information that can be effectively extracted by DL.
    \item Even if an attacker cannot use direction information because of distortions in patterns caused by a defense, she can still use timing information for an effective attack.
\end{smitemize}


\subsection{Directional Time}
\label{directionaltiming}

Since prior work~\cite{abe2016fingerprinting, Rimmer2018, Sirinam2018} has shown that packet direction is a powerful data representation for WF attacks, it should also be considered when using timing information. We evaluated different methods to combine timing and direction, such as using timing features and direction together, raw timing and direction together, and directional time. We find that directional time provides the most effective results. 

With the directional-time data representation, we experimented with WF attacks (again using 10-fold cross validation) and show the key results in Table~\ref{tab:CWresults}. Using directional time provides slightly higher accuracy than that of either using only direction or only raw timing in the Undefended dataset. Impressively, the attack against WTF-PAD can attain 93.4\% accuracy which is higher than that of either direction or raw timing. In the Onion Sites dataset, directional time has 12\% higher accuracy than using only direction information, slightly lower than using only raw timing information. Finally, directional-time performs very well against simulated W-T achieving 97\% accuracy and is able to completely undermine the defense.


\subsection{Real-world W-T Evaluation}

Table~\ref{tab:wt-real} presents attack performance when evaluated against our new real-world W-T dataset, which is described in Section~\ref{sec:real-wt-dataset}. We present this experiment separately from both the closed-world and open-world experiments, as the performance from this experiment cannot be compared in a straightforward manner. This dataset contains monitored and unmonitored site instances at an equal proportion (35K instances each). The goal of the classifier in this experiment is to both a) determine if the visited instance is from the monitored or the unmonitored set, and b) to determine which exact site was visited if belonging to the monitored set. To this end, we train our classifiers on data containing both the monitored and unmonitored instances, and we test on either exclusively monitored traces, exclusively unmonitored traces, or both together. If the defense works completely as intended, the combined test set should yield at most 50\% accuracy, as the classifier should confuse site pairs with one another.

\paragraphX{Results.} We see in Table~\ref{tab:wt-real} that, empirically, our W-T implementation does not provide the ideal results. The average accuracy for the combined test set is greater than 50\% for each of our three tested attacks. When examining our implementation and output data, we find that the cause of this shortcoming is the difficulty in exactly matching the target burst sequence when performing bust molding on the live network. Consequently, some information regarding the true class instance is leaked that allows the classifier to better distinguish between the true site and paired site. We discuss why our burst molding fails in more detail in Appendix~\ref{discussion:CW-W-T}.

In addition, we found that the training of the model was very unstable, with a high standard deviation between trials for all trace representations. This shows that while a practical implementation of W-T can not achieve ideal performance, the defense still significantly hampers the reliability of the classifier. We also see that direction and directional time representations achieve nearly identical results.

%
\begin{table}[tbp]
\renewcommand{\arraystretch}{1.25}
  \begin{center}
    \caption{{\em Open World:} Tuned for precision and tuned for recall. \textbf{D}: Direction, \textbf{RT}: Raw Timing, \textbf{DT}: Directional Time.}\vskip -0.2cm
    \label{tab:tunedprecionrecall}
    \begin{tabular}{llcccc}
    \multicolumn{2}{c}{\multirow{2}{*}{\textbf{Dataset}}}           & \multicolumn{2}{c}{\textbf{Tuned for Precision }} & \multicolumn{2}{c}{\textbf{Tuned for Recall}} \\ \cline{3-6} 
    \multicolumn{2}{c}{}                                                & \textbf{Precision}        & \textbf{Recall}       & \textbf{Precision}      & \textbf{Recall}     \\ \hline
    \multirow{3}{*}{\textbf{Undefended}} & \textbf{D}     & \textbf{0.991}                     & 0.938                 & \textbf{0.932}                   & \textbf{0.985}               \\
                                           & \textbf{RT}        & 0.969                     & 0.922                 & 0.857                   & \textbf{0.980}               \\
                                           & \textbf{DT} & \textbf{0.988}                     & \textbf{0.948}                 & 0.908                   & \textbf{0.989}               \\ \hline
    \multirow{3}{*}{\textbf{WTF-PAD}}      & \textbf{D}     & 0.961                     & 0.684                 & 0.667                   & \textbf{0.964}               \\
                                           & \textbf{RT}        & \textbf{0.972}                     & 0.609                 & 0.640                   & 0.942               \\
                                           & \textbf{DT} & \textbf{0.979}                     & \textbf{0.745}                 & \textbf{0.740}                   & \textbf{0.957}               \\ \hline
    \end{tabular}\vskip -0.6cm
   \end{center}
\end{table}
%

\subsection{Open-World Evaluation}

Having explored the quality of our models and made baseline comparisons of attacks in the closed-world setting, we now evaluate the attacks in the more realistic open-world setting. 

The performance of the attack is measured by the ability of the WF classifier to correctly recognize unknown network traffic as either a monitored or an unmonitored website. True positive rate (TPR) and false positive rate (FPR) have been commonly used in evaluating WF attacks and defenses in the open-world setting~\cite{wang2014effective, hayes2016k, Rimmer2018}. These metrics, however, could lead to inappropriate interpretation of the attacks' performance due to the heavy imbalance between the respective sizes of the monitored set and unmonitored set. Thus, as recommended by Panchenko et al.~\cite{panchenko2011website} and Juarez et al.~\cite{Juarez2014}, we use \emph{precision} and \emph{recall} as our primary metrics.


We trained the WF classifier by using the DF attack~\cite{Sirinam2018} as the base model with different data representations in both the Undefended and WTF-PAD datasets, including \emph{direction}, \emph{raw timing}, and \emph{directional time}. We did not evaluate the W-T defense in the open-world setting, as it remains a major challenge to obtain an open-world W-T dataset (see Appendix~\ref{discussion:OW-W-T}).

Finally, we note that the attacks can be flexibly tuned with respect to the attacker's goals. If the attacker's primary goal is to be highly confident that a user predicted to be visiting a monitored site truly is doing so, the attack should be tuned for precision, reducing false positives at the cost of also reducing true positives. On the other hand, if the attacker's goal is to widely detect any user that may be visiting a monitored website, the attack will be tuned for recall, increasing true positives while accepting more false positives. 

\begin{figure}[!t]
        \centering
        \hspace{-0.28cm}
%
%
%
\begin{tikzpicture}

\definecolor{color0}{rgb}{0.75,0,0.75}
\definecolor{color1}{rgb}{0.75,0.75,0}
\definecolor{color2}{rgb}{0,0.75,0.75}
\definecolor{color3}{rgb}{0.75,0,0}
\definecolor{color4}{rgb}{0,0.75,0}
\definecolor{color5}{rgb}{0,0,0.75}

\begin{axis}[
xlabel={Recall},
ylabel={Precision},
xmin=0.5, xmax=1,
ymin=0.5, ymax=1,
width=245,
height=170,
xtick={0.6,0.8,1},
ytick={0.6,0.8,1,1.2},
xmajorgrids,
x grid style={white!69.01960784313725!black},
ymajorgrids,
legend columns=1,
legend style={at={(0.51,0.57),font=\small}},
y grid style={white!69.01960784313725!black},
legend entries={{D-UnDef},{RT-UnDef},{DT-UnDef},{D-WTFPAD},{RT-WTFPAD},{DT-WTFPAD}}, legend cell align={left}
]

\addplot [thick, brown!50.0!black, dashed, mark=halfcircle*, mark size=\MarkerSize, mark options={solid},line width=\LineWidth,mark repeat={4}]
table {%
0.932164558223	0.985052631579
0.941283476503	0.980421052632
0.953765539916	0.977157894737
0.962820245782	0.973157894737
0.968497322272	0.970842105263
0.972718621127	0.968315789474
0.976481855911	0.965894736842
0.979954979097	0.962315789474
0.982966796033	0.959789473684
0.983987882722	0.957368421053
0.985649054142	0.954315789474
0.987430320254	0.950947368421
0.98910771262	0.946315789474
0.990592141671	0.942105263158
0.991101223582	0.937894736842
};

\addplot [thick, black, dashed, mark=diamond*, mark size=\MarkerSize, mark options={solid},line width=\LineWidth,mark repeat={4}]
table {%
0.857077	0.980316
0.860581	0.978526
0.875805	0.973158
0.895723	0.965684
0.91065	    0.961263
0.919717	0.957474
0.929047	0.953789
0.936889	0.948526
0.943148	0.944737
0.950521	0.940316
0.95557	    0.937263
0.959965	0.933895
0.963226	0.929158
0.966351	0.925053
0.969224	0.921579
};

\addplot [thick, red, dashed, mark=otimes*, mark size=\MarkerSize, mark options={solid},line width=\LineWidth,mark repeat={4}]
table {%
0.908291	0.989368
0.912091	0.986211
0.927493	0.982947
0.941409	0.979263
0.952572	0.976737
0.961235	0.973579
0.9674	    0.971474
0.970998	0.969158
0.974944	0.966632
0.977809	0.964737
0.980798	0.962421
0.982621	0.958211
0.985018	0.955053
0.98669	    0.952
0.987828	0.948211

};

\addplot [thick, color2, dashed, mark=square*, mark size=\MarkerSize, mark options={solid},line width=\LineWidth,mark repeat={2}]
table {%
0.66678800233	0.964105263158
0.68395042956	0.946947368421
0.734202947654	0.912421052632
0.785371364021	0.878210526316
0.819935366593	0.854631578947
0.846753524135	0.834631578947
0.870096390944	0.817157894737
0.891319606926	0.802
0.905582524272	0.785473684211
0.918905347728	0.770526315789
0.930680192483	0.753263157895
0.940796555436	0.736
0.947675225538	0.718736842105
0.953938872148	0.699789473684
0.960798816568	0.683684210526
};

\addplot [thick, green!50.0!black, dashed, mark=triangle*, mark size=\MarkerSize, mark options={solid,rotate=180},line width=\LineWidth,mark repeat={2}]
table {%
0.640103	0.941895
0.664258	0.909684
0.730715	0.865474
0.788477	0.828316
0.828205	0.801789
0.859199	0.779158
0.882981	0.756947
0.903799	0.738737
0.918088	0.719684
0.930336	0.701474
0.94235	    0.681368
0.950884	0.662316
0.959117	0.644526
0.965512	0.627684
0.971798	0.609368
};

\addplot [thick, purple, dashed, mark=star, mark size=4, mark options={solid},line width=\LineWidth,mark repeat={2}]
table {%
0.739671	0.957368
0.750665	0.950737
0.796603	0.928
0.842043	0.905684
0.870256	0.888211
0.888948	0.872947
0.909293	0.86
0.923713	0.846316
0.935683	0.83
0.946084	0.816421
0.955478	0.804211
0.963996	0.789158
0.968766	0.773789
0.974314	0.758632
0.979259	0.745474

};
\end{axis}

\end{tikzpicture}
        \vspace{-0.2cm}
        \caption{{\em Open World:} Precision-Recall curves. Note that both axes are shown for 0.5 and above.}\label{fig:PreRec_OW}
\end{figure}
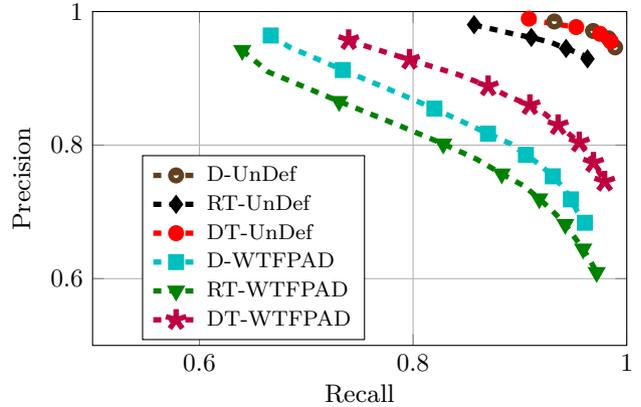

\begin{table*}[tbp]
\renewcommand{\arraystretch}{1.25}
  \begin{center}
    \caption{{\bf Closed World:} Evaluation of Slowest \& Fastest Circuits.\\ 
    \textbf{D}: Direction, \textbf{RT}: Raw Timing, \textbf{DT}: Directional Time.}\vskip -0.2cm
    \label{tab:congestion}
    \begin{tabular}{lccc|ccc}

    \multicolumn{1}{c}{\multirow{2}{*}{\textbf{Dataset}}}           & \multicolumn{3}{c|}{\textbf{Slowest Circuits as Test Set}} & \multicolumn{3}{c}{\textbf{Fastest Circuits as Test Set}} \\ \cline{2-7}
    
    
        &   \textbf{D}~\cite{Sirinam2018} & \textbf{RT}       & \textbf{DT}       & \textbf{D}~\cite{Sirinam2018} & \textbf{RT}       & \textbf{DT}     \\ \hline
    
\textbf{Undefended}     & \textbf{92.90\%}   & 82.30\%  & 92.20\%    & 93.80\%    & 82.50\%  & \textbf{94.40\%}  \\
\textbf{WTF-PAD} & 68.40\%  & 42.80\%  & \textbf{71.10\%}    & 69.20\%   &  44.20\%  & \textbf{74.10\%}  \\
\textbf{Onion Sites} & 47.20\% & 52.90\% & \textbf{53.50\%} & 41.30\% & \textbf{47.10\%} & 40.60\% \\
     \end{tabular}\vskip -0.8cm
   \end{center}
\end{table*}

\paragraphX{Results.}
Figure~\ref{fig:PreRec_OW} shows precision-recall curves for the attacks in the open-world setting, while Table~\ref{tab:tunedprecionrecall} shows the results when the attack is tuned for precision or tuned for recall. For the Undefended datasets, the results show that all data representations can effectively be used to attain high precision and recall. The attacks consistently performed best on the \emph{direction} and \emph{directional time} data representations, with 0.99 precision and 0.94 recall when tuned for precision. Timing alone, however, is also very effective. 

On all three WTF-PAD datasets, we see reductions in both precision and recall. Nevertheless, all three datasets show over 0.96 precision and 0.60 recall when tuned for precision. Interestingly, Figure~\ref{fig:PreRec_OW} shows that \emph{directional time} outperforms \emph{direction} on WTF-PAD data. Timing information appears to improve classification of monitored versus unmonitored sites for traffic defended with WTF-PAD. 

\subsection{Impact of Congestion on Circuits}
\label{congestion_circuits}
\begin{figure}[!t]
        \centering \vskip -0.4cm
        \includegraphics[scale=0.50]{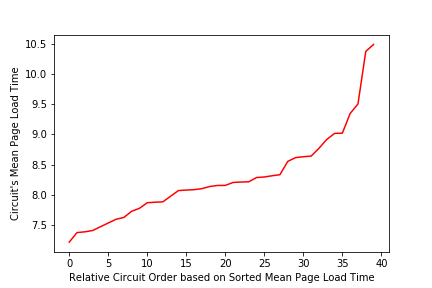}
        \caption{Distribution of mean circuit page load time for site 0 (Undefended dataset). Y-axis starts from 7.0.}
        \label{circuit_site_0}
        \vskip -0.2cm
\end{figure}

Even though we split training, validation, and testing sets based on circuits as described in Section~\ref{split_circuits}, there can be situations where the victim has a much different circuit than the ones used by the attacker for training the classifier. In particular, circuit congestion might be expected to especially impact timing features. To investigate this, we perform two types of experiments: i) the slowest 10\% of circuits as the testing set, and ii) the fastest 10\% of circuits as the testing set. These experiments model the scenario in which the attacker trains on a broad range of circuits (the other 90\% of the data), but the victim has an unusually slow or fast circuit, respectively. To split off the slower and faster circuits for a given monitored site, we rank the circuits based on the mean page load time among the page load times for that site. The mean page load time (in seconds) of the fastest and slowest four circuits of site 0 in our Undefended dataset are [7.21, 7.37, 7.38, 7.40] and [9.34, 9.50, 10.37, 10.48], respectively (see Figure~\ref{circuit_site_0} for the whole distribution). 

Among all sites in the Undefended dataset, the median gap in page load time between the fastest and slowest circuits for the site was 5.11s (15.79s slow and 10.68s fast), showing a substantial difference in the typical case. For the Onion Site dataset, the median gap was much higher, at 15.69 (20.32s slow and 4.63s fast). Some Onion Site circuits can be very quick, likely due to not needing to use exit nodes, which can be highly congested.




We perform these experiments with Undefended, WTF-PAD, 
and Onion Sites datasets. The Undefended and WTF-PAD datasets each contain 40 circuits with 25 visits in each circuit. We take the data from the slowest or fastest four circuits as the test set and evenly split the rest of the 36 circuits into training and validation sets. For 
the Onion Sites, each visit is from a different circuit, so we take the 10\% of the slowest or fastest circuits as the test set and the rest of the 90\% circuits as the training and validation sets.

 
\paragraphX{Results.} As Table~\ref{tab:congestion} shows, attack accuracies are lower in all scenarios for all datasets. Interestingly, \textit{direction} is impacted almost as much as \textit{raw timing} and \textit{directional time} attacks (and in some cases more so), indicating that differences in circuit speed impact burst patterns significantly, not just timing patterns. In most cases, testing with the slowest circuits harms accuracy less than training with the fastest circuits. The exception is for the Onion Sites, where the fastest circuits can be much faster than slower ones, leading to confusion for the attacker.
\section{Discussion}\label{discussion}


In this section, we discuss the reasons why WTF-PAD and W-T have their respective levels of vulnerability to leaking timing information.

\paragraphX{WTF-PAD Defense.} To better understand why timing information improves attack performance in WTF-PAD, we further examined the information leakage of timing features, as shown in Table~\ref{tab:jointleak}. Leakage is reduced for all feature groups except the Packet per Second category, but some feature categories like MED have minimal reduction.
This indicates that the classifier is still able to find valuable information in the timestamps of the real bursts. This intuitively makes sense as the timestamps of real bursts are unaffected by WTF-PAD's zero-delay adaptive padding. The classifier thus needs to only learn to distinguish between the timings of real and fake bursts to improve classification performance.


\paragraphX{W-T.} In Section~\ref{SecExperiments}, we showed that W-T leaks a large amount of information from packet timestamps in the simulated setting, leading to high classification accuracies. This is because the simulated dataset can mold bursts without affecting timestamps of real packets, leaving substantial timing information unchanged. 
We found, however, a discrepancy in performance between our simulated and real-world testing. This is likely due to the compromises that needed to be made when practically implementing burst-molding in Tor. A side-effect of our implementation is that inter-burst packet timing variance is virtually eliminated and intra-burst timing variance is reduced (see Appendix~\ref{sec:wtprototypedesign} for implementation details). Consequently, the value of timing information is significantly lessened. We also note, however, that as a consequence of difficulties in implementing burst molding, directional information is leaked which has led to improved performance of directional information against our prototype.

\section{Conclusion}
In this study, we investigated the use of timing information as a source of features to perform effective WF attacks on Tor. 
We proposed eight new burst-level timing features that help illustrate how timing can be reliably used to fingerprint sites. 
Through experiments with machine-learning and deep-learning classifiers, we show that these features are robust over multiple noisy instances and provide meaningful classification power. Furthermore, we show that these features have low redundancy with previous studied features based on timing.



Since we found that a range of timing features are relevant for website fingerprinting, we then explored the capability of deep-learning classifiers to extract and use features such as these. First, we show that based only on timestamps and ignoring packet direction, an attacker can achieve surprisingly good results, such as 96\% accuracy on the Undefended dataset. 
Moreover, we proposed the use of directional timing, formulated by taking the product of timing and direction data, and we show that this improves the performance of the attack over using just direction or just timing in most settings in the closed world. For example, on Onion Sites, directional timing is 12\% more accurate than direction alone.

In more realistic open-world experiments on undefended traffic, using directional timing information, the attack attains 0.99 precision and 0.95 recall. Against WTF-PAD, it reaches 0.98 precision and 0.75 recall. These are 
modest improvements in attack performance when compared to using only directional information. 

In summary, our study shows that timing information can be used as an additional input to create an effective WF attack.
Furthermore, these findings show that developers of WF defenses need to pay more attention to timing features as another fingerprintable attribute of the traffic. Timing information seems to affect defenses in ways that are hard to predict, so both evaluating with our attacks and performing feature-based leakage analysis are important steps in understanding if a defense is sound.

\paragraphX{Acknowledgements.}
We thank the anonymous reviewers for their helpful feedback. We give special thanks to Tao Wang for providing details about the technical implementation of the W-T defense, and to Marc Juarez for providing guidelines on developing the W-T prototype. This material is based upon work supported in part by the National Science Foundation under Grants No.\ 1722743 and 1816851.

{\normalsize \bibliographystyle{acm}
\bibliography{main}}

\appendix

\section{W-T Data Collection}\label{sec:wtdata}

In order to accurately evaluate a defense against our timing-based attack, we require a dataset that contains realistic timestamp information. At the time the Walkie-Talkie (W-T) defense was proposed, Wang et. al. did not consider attacks using timing as a credible threat against Tor traffic~\cite{wang2017walkie}. As such, their defense simulator did not calculate the timestamps of dummy packets in a realistic manner. To address this gap, we have developed a prototype of the W-T defense specification that runs directly on the Tor network, rather than simulating the padding.

\begin{figure}[tbp]
	\centering
    \includegraphics[scale=0.20]{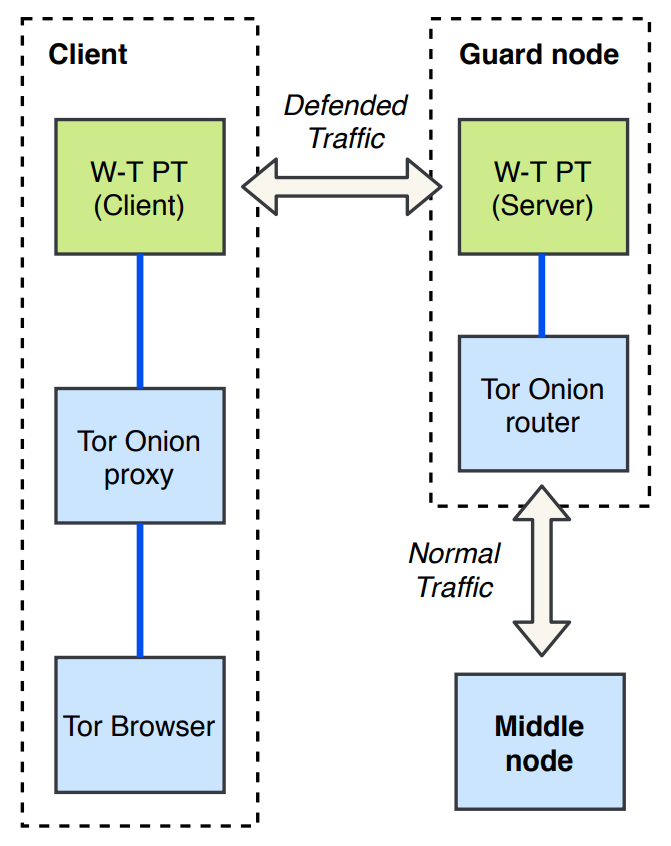}
    \caption{Walkie-Talkie Prototype Deployment.}
    \label{transport}
    \vskip -0.3cm
\end{figure}

The W-T prototype is designed as a Tor Pluggable Transport (PT) module, as an extension of the WFPadTools Framework~\cite{wfpadtools}. We developed our own implementation of the W-T padding algorithm to use in this PT. The PT is deployed on both the client and the guard node.\footnote{The WFPadTools Framework operates on a Tor bridge node, but we have the bridge act as a guard so as to not add a node to the length of the circuit.} Figure~\ref{transport} shows how the W-T PT operates in the context of the Tor network. Our prototype is intended to be used in tandem with the Tor Browser Bundle configured with the half-duplex patch used in Sirinam et al.'s evaluation of W-T~\cite{Sirinam2018}.

It is important to note that we have deployed our defense prototype on the Tor Guard node since we are primarily interested in evaluating the performance of the padding mechanism. This deployment is vulnerable to a malicious guard node. A practical deployment would ideally be done on the middle node so that the security of the system is not so heavily dependent on the trustworthiness of the Guard. However, this style of deployment could not be done through a PT and would instead need to be implemented directly into the Tor network code.

\subsection{W-T Dataset}

Using our W-T prototype, we performed a new crawl of 100 monitored websites to collect a new W-T dataset. The W-T defense creates pairs between two or more different websites and performs padding to make the two sites look the same in terms of their packet sequence. Wang and Goldberg's dataset keeps the pairings the same for the full dataset. Unfortunately, this does not accurately model the attack. While a client should indeed use the same pairings all the time, the attacker should not know what those pairings are in advance.

Consequently, an attacker cannot train on a dataset that contains only the correct pairings. The attacker must instead train on many possible pairings of real and decoy sites. With this understanding, we designed our real-world crawl. We first sampled a list of monitored and unmonitored websites from the Alexa Top websites. We use the top 100 sites for our monitored sites and sample 10,000 sites from the top 14,000 for our unmonitored set. We next randomly generated pairings between the monitored and unmonitored sites such that each new sample to be collected is composed of a new pair. When we collect our samples, we generate one visit for both the sites in each pair such that samples for both the real site sequence and decoy site sequence are represented. Based on the W-T defense, a client chooses the decoy site. In a realistic setting, the paired traffic between the actual website and the decoy can be variously different among different users. Hence, an attacker needs to train the classifier with traffic paired with variety of other sites. In this way an attacker can test on W-T traffic of any user.

Our crawl was performed using a modified variant of the Tor Browser Crawler~\cite{torcrawler} so as to accurately represent the browsing behavior of the Tor Browser Bundle. The crawlers were configured to use version 7.0.6 of the Tor Browser Bundle, patched so as to operate in half-duplex mode. We deployed our W-T defense prototype as Tor guards on virtual private servers hosted by Amazon Web Services. Our crawlers were then configured to connect to these servers as their guards. We collected our data in batches in which each site was visited once. Between each batch, the decoy site to which each real site was paired was changed, following the pairing scheme discussed previously.

\section{W-T Prototype Design}
\label{sec:wtprototypedesign}

Designing and developing an experimental prototype of W-T led us to face many issues that are important to understand when designing WF defenses for Tor, and we address these in this section.

\paragraphX{Burst Identification.}
Implementing W-T padding requires the defense to know which burst the stream is currently on so it knows how much padding is required. So it is necessary to correctly identify when a new W-T burst had begun. It is however difficult for the co-operating Tor node to know when the current burst coming from upstream has ended. We solve this by implementing a half-duplex communication mechanism in the PT. The PT allows only one side to send data at any given time. Additionally, time thresholding is used to identify when the current burst has ended. If no packets are seen after a certain amount of time, the current burst is determined to be over. This information is then signalled to the other side of the connection by piggybacking a control message onto the last message in the burst. These mechanisms allow both the client PT and guard PT to remain synced to the current position in the burst sequence.

\paragraphX{Padding.}
The next decision to be made is when to send the dummy packets in burst. If we send the dummy packets at the start of the burst as Wang and Goldberg describe in~\cite{wang2017walkie}, we must assume knowledge of the real burst beforehand. In practice, we found this results in many errors as the true burst sequence changes between visits. Instead, we opt to send data in a burst all at once by keeping outgoing packets in a queue until the burst ends. This provides two benefits: 1) the number of necessary dummy packets to add can be accurately computed using the true burst size, and 2) the inter-burst packet timings become nearly identical and the authenticity of a packet cannot be distinguished by timing discrepancies. 

\paragraphX{Tail Padding.}
Fake bursts need to be added to the trace when the real burst sequence is shorter than the burst sequence of the decoy site. The W-T specification gives no guidelines as to when in the burst sequence to add these bursts. For our implementation, we simply add the fake bursts at the end of the real communication, which we identify based on when the Tor browser closes its connection to the proxy application.

\subsection{Limitations of the Implementation}\label{discussion:W-T}

\paragraphX{Burst Identification.}
Our burst identification process works well so long as packets in a burst do not have timing differences greater than the threshold. Unfortunately, very large timing threshold cannot be selected without also inducing additional an additional latency overhead. Furthermore, if the time threshold is too small, the threshold may expire before any packets in the next burst have arrived. This is most likely to occur on the co-operating Tor node as there may be several hops worth of distance for packets to travel before reaching and arriving from the end website. These problems can be reduced by configuring different time thresholds for different scenarios and tuning (eg. we don't allow the burst to end if packets have yet to arrive unless a much larger time threshold expires). However, discrepancies will always exist and the PT will occasionally incorrectly end bursts. When this happens, the traffic may be segmented into more bursts than necessary which can result in the trailing portion of the traffic to receive less padding than it otherwise would if decoy sequence is smaller than the real site.

\paragraphX{Overheads.}
The overhead of our W-T implementation is heavily dependent on two factors: (1) the burst segmentation time threshold, and (2) the scheme used for pairing sensitive and non-sensitive sites. For our data collection, we used a large time threshold of 300ms in order to minimize the number of occurrences in which a burst is ended early. Furthermore, we made no attempt to optimally pair sites of similar lengths. As a consequence of this, we see high bandwidth and latency overheads when compared to their reference sequences. When find that packet overheads average $2.21\pm1.22$ times the original sequence and time overheads average $10.13\pm6.48$ times. In practice, these overheads likely can be reduced by more optimally pairing similarly sized traces and by tuning the time threshold to the minimum value that still segments burst with reasonable accuracy.

\paragraphX{Padding.}
In Appendix~\ref{sec:wtdata} we presented a padding scheme that allows our prototype to reliably manipulate a burst sequence to match a target sequence by computing the necessary number of padding packets on the fly. This scheme however has a limitation. Dummy packets can only be added to increase the size of a burst, real packets cannot be dropped without causing communication errors. Consequently, in instances where dynamic content or burst identification errors yield larger than expected burst sizes, the burst size cannot be manipulated to match a smaller target size in this scheme. This will inevitably result in leaks of information that a classifier may use to better distinguish between real and decoy sites.

\paragraphX{Tail Padding.}
As described in Appendix~\ref{sec:wtdata}, the W-T specification does not indicate at what point the defense should create a fake burst. The difficulty of adding a fake burst is that the timing of the packets within the burst and between bursts should resemble that of real bursts. Otherwise, the attacker needs only to identify and filter out likely fake bursts to greatly improve their classifier's performance. This issue is magnified if the fake bursts are left until the real traffic ends, as done in our implementation. If the attacker can identify one fake burst the attacker can prune the trace to the last suspected real burst.

\section{W-T Experimental Results}\label{discussion:CW-W-T}
The experimental results show that W-T is in practice weaker than the 50\% maximum attacker accuracy it claims to guarantee~\cite{wang2017walkie}. The cause of this discrepancy is the defender's inability to perfectly manipulate the traffic they produce to match their target pairing.

We note that both the original W-T dataset created by Wang and Goldberg and the dataset developed by Sirinam et al. used traces of half-duplex network traffic and simulated the padding~\cite{wang2017walkie, Sirinam2018}. Given this controlled and simulated condition, the traffic from the two websites can be formed into a supersequence via an algorithm that is strictly followed without dealing with other factors, such as the usual variance in burst sequences between site loads, the effect of padding on the network connection, and the processing time on the Tor nodes. 

In contrast, our W-T dataset was directly crawled from our W-T prototype, which was built to work with padding on the Tor network. This not only allowed us to evaluate W-T with realistic timestamps, but also uncovered the issues discussed in Appendix~\ref{discussion:W-T}. The instance-to-instance changes we observed in the traces would have led to different burst sequences than expected by the algorithm, exposing fingerprints that could be detected by the DF classifier. We note that W-T still maintains a lower classification accuracy than WTF-PAD when tested against our attacks, so it appears that the supersequence-based padding may still be more desirable.

The effects we find from realistic conditions raise the questions about the actual performance of defenses. While simulated padding is useful for gathering an initial idea about a defense's effects, padding should be evaluated experimentally before confident judgments can be made about its design. 

\section{Open World Challenge of W-T}\label{discussion:OW-W-T}
The fundamental concept of W-T in the open-world setting is to attempt to confuse the classifier by creating a supersequence between a monitored website and an unmonitored website~\cite{wang2017walkie}. This simple idea, however, is not easily implemented nor tested for a few main reasons:
\begin{itemize}
    \item Since each attacker selects his own monitored set, we cannot expect to know what the monitored sites are. Supposing that the W-T algorithm pairs some sites that are more likely to be monitored (sensitive sites) with those that are less likely (non-sensitive sites), one must still test how attacks perform when the attacker monitors sensitive sites, non-sensitive sites, and a mix of both.
    
    \item Each user may have a different idea of what sites are sensitive and not sensitive and should be paired together. If pairing is random, a sensitive site might be paired with a particularly unlikely decoy site, greatly reducing its effective protection. So finding a good approach to pairing sites is an open challenge. 

    \item When W-T pairs a real site $A$ and an decoy site $B$, this pairing must be kept the same for all future visits of site $A$. Otherwise, the attacker will see site $A$ paired with different sites and can eventually infer that $A$ is being visited, while the other sites are decoys. Further, the pairing must be symmetric, such that if the user actually visited the decoy site $B$, the site $A$ must be selected as its decoy. This could be achieved by locally storing the mapping of decoy and real sites, but this would need to be done carefully to avoid leaving a record of the user's Tor activity on the client. Alternatively, every possible site the user could visit could be paired up in advance, but this an enormous list of sites. Note that W-T also requires a database of traffic traces for every possible site, so it is already a problem before pairing is considered.
    
    
    
    
\end{itemize}
These issues must be carefully addressed before a realistic study of W-T in the open-world setting could be conducted. Furthermore, the issues with the site-pairing algorithm remain major problems to address before W-T could be deployed.

\end{document}